\documentclass[structabstract]{aa}  
\usepackage{graphicx}
\usepackage{txfonts}
\begin{document}

  \title{Long-slit spectrophotometry of the multiple\\ knots of the polar ring galaxy IIZw71}
\titlerunning{Long-slit spectrophotometry of IIZw71}

  \author{E. P\'erez-Montero\inst{1,2}
 \and R. Garc\'\i a-Benito \inst{3}
 \and A.I. D\'\i az \inst{3}
 \and E. P\'erez \inst{1}
\and C. Kehrig \inst{4}}

 \offprints{E. P\'erez-Montero}

 \institute{
 Instituto de Astrof\'\i sica de Andaluc\'\i a (CSIC)
        Apartado de Correos 3004. 18080,  Granada, Spain.\\
             \email{epm@iaa.es, eperez@iaa.es}
 \and
 Laboratoire d'Astrophysique de Toulouse et Tarbes, Universit\'e de Toulouse, CNRS, 14 avenue E. Belin, 31400 Toulouse, France
              \email{}
\and
Departamento de F\'\i sica Te\'orica, 
         C-XI, Universidad Aut\'onoma de Madrid,
              28049, Cantoblanco, Madrid, Spain \\
 \email{ruben.benito@uam.es,angeles.diaz@uam.es}
      \and
                     University of Michigan, Department of Astronomy, 830 Dennison Building, Ann Arbor, MI 48109-1042\\
             \email{kehrig@umich.edu}
}

   \date{}


%

   \abstract
{}
{The blue compact dwarf galaxy IIZw71 is catalogued as a probable polar-ring galaxy, and along its long axis it has several
very luminous knots showing recent episodes of star formation. Our main aim is to study the physical properties, the stellar content, and the kinematics in the brightest knots of the polar ring}
{We carried out long-slit 
spectroscopic observations of the polar ring in the spectral range 3500 - 10000 {\AA} taken with the William Herschel Telescope (WHT).  The spectroscopic observations complemented by the available photometry of the galaxy in narrow H$\alpha$ filters.}
{ We measured the rotation curve of the ring, from which we infer a ratio $M/L_B \approx$ 3.9 inside the star-forming ring. 
We measured the auroral [O{\sc iii}] line in the two brightest knots, allowing us to measure oxygen, sulphur, nitrogen, argon, and neon chemical abundances  following the direct method. Different empirical calibrators were used to estimate the oxygen abundance in the two faintest knots where the temperature sensitive lines could not be measured. The metallicities obtained are very similar for all the knots, but lower than previously reported in the literature from integrated spectra. The N/O abundance, as derived from the N$_2$O$_2$ parameter (the ratio of the [N{\sc ii}] and [O{\sc ii}] intensities), is remarkably constant over the ring, indicating that local polution processes are not conspicuous.
Using synthetic stellar populations (SSPs) calculated with the code STARLIGHT, we studied the age distribution of the stellar populations in each knot, finding that in all of them there is a combination of a very young population with less than 10 Myr, responsible for the ionisation of the gas, with other populations older than 100 Myr, probably responsible for the chemical evolution of the knots. The small differences in metallicity and the age distributions among the different knots are  indicative of a common chemical evolution, probably related to the process of interaction with the companion galaxy IIZw70. Finally, we calculated the SFR in the different knots from the H$\alpha$ luminosities. The combined SFR rate for the ring amounts to half of the integrated SFR for this galaxy reported by previous authors. The average surface SFR is also higher than the average reported value of HII in polar rings by more than an order of magnitude.}
{} 
\maketitle

%

\section{Introduction}

The intense processes of star formation occurring in the luminous knots of
HII galaxies make them easily detected in surveys based on strong
emission lines. In fact, these knots present spectra similar to those of 
HII regions, so it is possible to use them to ascertain the basic properties 
of the host galaxy, such as extinction, metallicity, star formation rates, 
and from them derive the chemical evolution and the star formation history of 
the whole galaxy.

HII galaxies are a subset of blue compact dwarf galaxies (BCDs), which are
selected by their compact aspect and blue colour.
This type of object is thought to be more frequent in a younger Universe and 
they are possible building blocks for the largest galaxies that we can detect at 
low redshifts (Kauffmann et al., 1993). If interactions among dwarf irregular galaxies was a basic 
mechanism of galaxy formation in the past, it is important to study the cases 
taking place in the nearby Universe now and the links between dwarf interactions and the
star formation history of BCDs.

Although IIZw71 is catalogued as a BCD galaxy it is, in fact, characterised by several very 
luminous H$\alpha$ knots distributed along a ring that is rotating around the host galaxy. It has been catalogued as a
probable polar-ring galaxy (B17) in the Polar-Ring Catalogue (Whitmore et al., 1990). 
Polar ring galaxies (PRGs) are systems with two kinematically separated components. The central component (the host galaxy)
is usually a lenticular galaxy or occasionally an elliptical galaxy. The other component, the polar ring, follows a
approximately circumpolar orbit around the host and it is characterised by the presence of stars, molecular gas ,and dust, inside another bigger ring composed of neutral hydrogen. Thus, this ring becomes an appropriate place for star formation.
It is thought that these objects are formed as a consequence of the
interaction between galaxies with a small impact parameter (Bournaud \& Combes, 2003). In the case of IIZw71, there are proofs from
interferometric observations of interaction with IIZw70, another BCD. In fact, although clearly separated on optical images, both galaxies share a common HI envelope with a gaseous bridge or streamer connecting both structures (see Figure 3 of Cox et al., 2001). This points to an ongoing interaction between the two galaxies.
Therefore this system becomes an ideal scenario for studying the effects of interactions in the formation and evolution of BCDs in the Local Universe. A distance of 18.1 Mpc to the system is adopted by Cox et al. (2001), taking a value of H$_o$ = 75 Km s$^{-1}$ Mpc$^{-1}$. This implies a linear scale of 90 pc/arcsec on the sky.

In a single long-slit exposition, we have observed the main bursts of star formation along the direction of the polar ring of IIZw71.
The analysis of the obtained spectra and the available 
photometry in several bands were used to make a comparative analysis of 
the different bursts of star formation. In the next section we describe the long-slit spectroscopic
observations of the polar ring of IIZw71.  In sect. 3, we present the results of our study, including
the determination of physical properties, such as electron density and reddening in the four observed knots,
and of chemical abundances in the two brightest knots. In sect. 4, we discuss 
our results including a study of the kinematics of the polar ring, the determination of metallicity in both
the faintest and the brightest knots, the reddening,
the stellar properties by means of fitting synthesis stellar populations. and the star formation rates in
the individual knots using the available H$\alpha$ photometry of this galaxy.
Finally, our conclusions are presented in the last section.

\section{Observations and reduction}


The long-slit spectrophotometric observations of IIZw71 were obtained using the ISIS double-beam spectrograph mounted on the 4.2m William Herschel Telescope (WHT) of the Isaac Newton Group
(ING) at the Roque de los Muchachos Observatory on the Spanish island of La
Palma. They were acquired on 2005 July 8 during one 
single night's observing run and under photometric conditions, with an average seeing of 0.7 arcsec.
The EEV12 and Marconi2 detectors were attached to the blue and red
arms of the spectrograph, respectively. The R600B grating was used in the blue covering the
wavelength range 3670-5070\,\AA\ (centred at $\lambda_c$\,=\,4370\,\AA), giving
a spectral dispersion of 0.45\,\AA\,pixel$^{-1}$. On the red arm, the R316R
grating was mounted in two different central wavelengths providing a spectral range from 5500 to
7800\,\AA\ ($\lambda_c$\,=\,6650\,\AA) and from 7600 to 9900 \,\AA\ ($\lambda_c$\,=\,8750\,\AA) 
with a spectral dispersion of 0.86\,\AA\,pixel$^{-1}$. To reduce the readout noise of our images, the observations were taken with the `SLOW' CCD speed. The pixel size for this set-up
configuration is 0.2 arcsec for both spectral ranges.  The slit width
was $\sim$1 arcsec, which, combined with the spectral dispersions, yields
spectral resolutions of about 1.0 and 3.5\,\AA\ FWHM in the blue and red arms, 
respectively. The instrumental configuration, summarized in
Table \ref{config}, was planned to cover the
whole spectrum from about 3500 to 10000\,\AA\  , at the same time providing a moderate
spectral resolution. This guarantees the simultaneous measurement 
of the nebular lines of [O{\sc ii}]\,$\lambda\lambda$\,3727, 3729,
and [S{\sc iii}]\,$\lambda\lambda$\,9069,9532\,\AA\  at both ends of the
spectrum, in the very same region of the galaxy. 


\begin{table}
\centering
\caption[]{WHT instrumental configuration}
\label{config}
\begin{tabular} {l c c c c c}
\hline
Spectral range  &       Disp.             & FWHM & Spatial res.    & Exposure Time    \\
    (\AA)             & (\AA\,px$^{-1}$) & (\AA)    & (\arcsec\,px$^{-1}$) & s \\
\hline
3670-5070   &       0.45              &  1.0      &   0.2   & 4$\times$ 1200            \\
5500-7800  &       0.86             &  3.5      &   0.2     & 2$\times$ 1200        \\
7600-9900  &       0.86             &  3.5      &   0.2     & 2$\times$ 1200          \\
\hline
\end{tabular}
\end{table}


To cover all knots along the polar ring at the same exposure, all the spectra were taken at PA= 30$^{\rm o}$, an average of 56$^{\rm o}$ off paralactic angle, and the spectra are thus somehow
affected by atmospheric differential refraction. Taking the mean
air mass into account during the observations of this object ($\approx$ 1.25) and following
the curves by Filippenko (1982), we calculated that the displacement
between [O{\sc ii}] 3727 {\AA} and [S{\sc iii}] 9069 {\AA}, which are the emission lines the most separated 
in wavelength, is no greater than 0.9 arcsecs.

Several bias and sky flat field frames were taken at the beginning and at the end
of the night in both arms. In addition, two lamp flat fields and one calibration
lamp exposure were performed at each telescope position. The calibration lamp
used was CuNe+CuAr. The images were processed and analysed with
IRAF\footnote{IRAF: the Image Reduction and Analysis Facility is distributed by
  the National Optical Astronomy Observatories, which is operated by the
  Association of Universities for Research in Astronomy, In. (AURA) under
  cooperative agreement with the National Science Foundation (NSF).} routines in 
the usual manner. This procedure includes
the removal of cosmic rays, bias subtraction, division by a normalised flat
field, and wavelength calibration. Typical wavelength fits were performed using 30-35
lines in the blue and 20-25 lines in the red and polynomials of second to
third order. These fits were made at 117 different locations along the slit
in each arm (beam size of 10 pixels) obtaining rms residuals between
$\sim$0.1 and $\sim$0.2\,pix.  
In the last step, the spectra were corrected for atmospheric extinction and flux-calibrated. 
For both arms, two standard star observations were used, allowing a good
spectrophotometric calibration with an estimated accuracy of about 5\%. 

\begin{figure*}[t]
\centering
   \includegraphics[viewport = 130 0 510 800, width=8.5cm,angle=270,clip=]{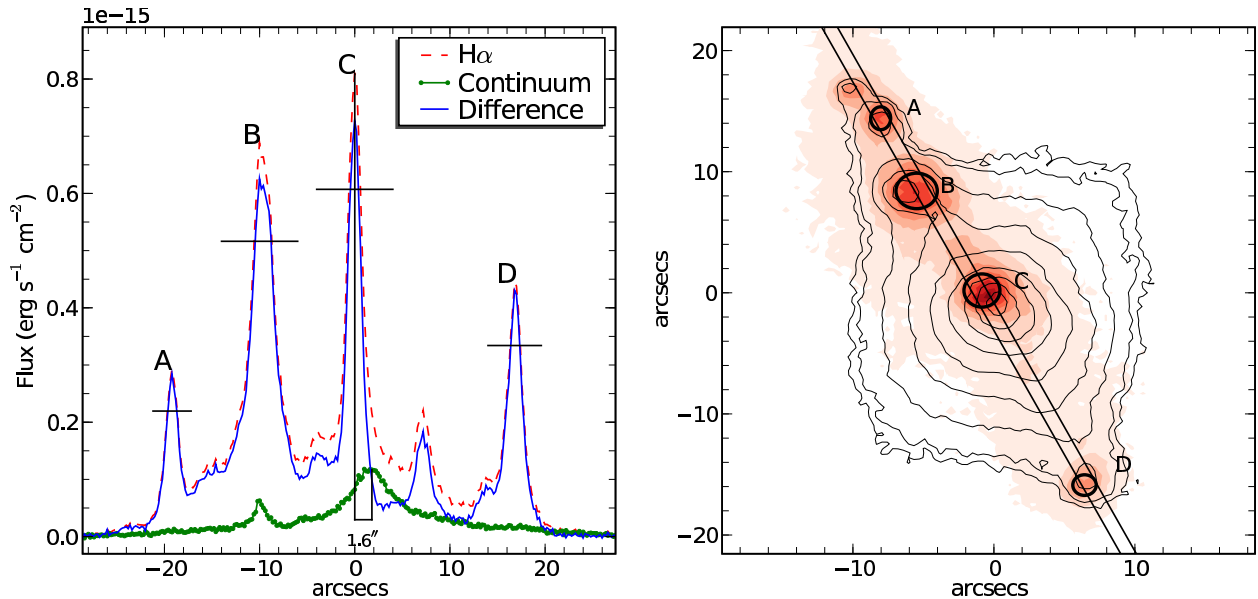}
   \caption{In left panel, spatial profile of the light distribution along the slit for the observed H$\alpha$ emission (dashed line), the adjacent continuum ({\bf thick green line}), and the difference between the two ({\bf thin blue line}). The horizontal lines show the spatial pixels that have been compressed to produce the one dimensional spectra corresponding to each emission knot. In the right panel, we show the H$\alpha$ image from Gil de Paz et al. (2003) with the identification of the observed knots and the regions for which the broad band magnitudes and the narrow H$\alpha$ flux have been measured. In the same image, we show the position of the slit for the observations described in the text and the elliptical regions measured to correct for aperture effects in each knot. We show the R-band contours as well, as taken from the same source,
which show the position of the host galaxy.}
              \label{profile}
    \end{figure*}

The left panel of Figure \ref{profile} shows the spatial distribution of the H$\alpha$ flux along the slit.  
The emission line profile in this plot has been generated by collapsing 11 pixels of the spectra in the
direction of the resolution at the central position of the line in the rest
frame, $\lambda$\,6563 \AA. A continuum profile was also generated by collapsing 11 resolution
pixels centred at $\lambda$\,6593\,\AA\ . The difference between the two, corresponds
to the pure H$\alpha$ emission. 
It can be see from the profiles that in knot C, the most intense one, the peaks of the continuum and line 
emission do not coincide spatially but are separated by 1.6 arcsec. 
Also the width of the continuum profile is larger than that of the line emission. 
This suggests that most of this continuum is coming from the host galaxy behind the knot.

The regions of the frames to be extracted into one-dimensional spectra corresponding to each of the knots 
were selected on the H$\alpha$ line and are marked by horizontal lines in the left panel of Figure \ref{profile}.
In the right panel of the same figure we show the H$\alpha$  image and continuum contours in the R-band 
from Gil de Paz et al. (2003) in order to illustrate the position of the bursts of star formation and the 
relative position of the host galaxy. We also show elliptical regions in H$\alpha$ taken to measure 
the total H$\alpha$ flux of the knots for an aperture correction of the slit measurements. 
The different emission knots are labelled in both panels.

\begin{figure*}[t]
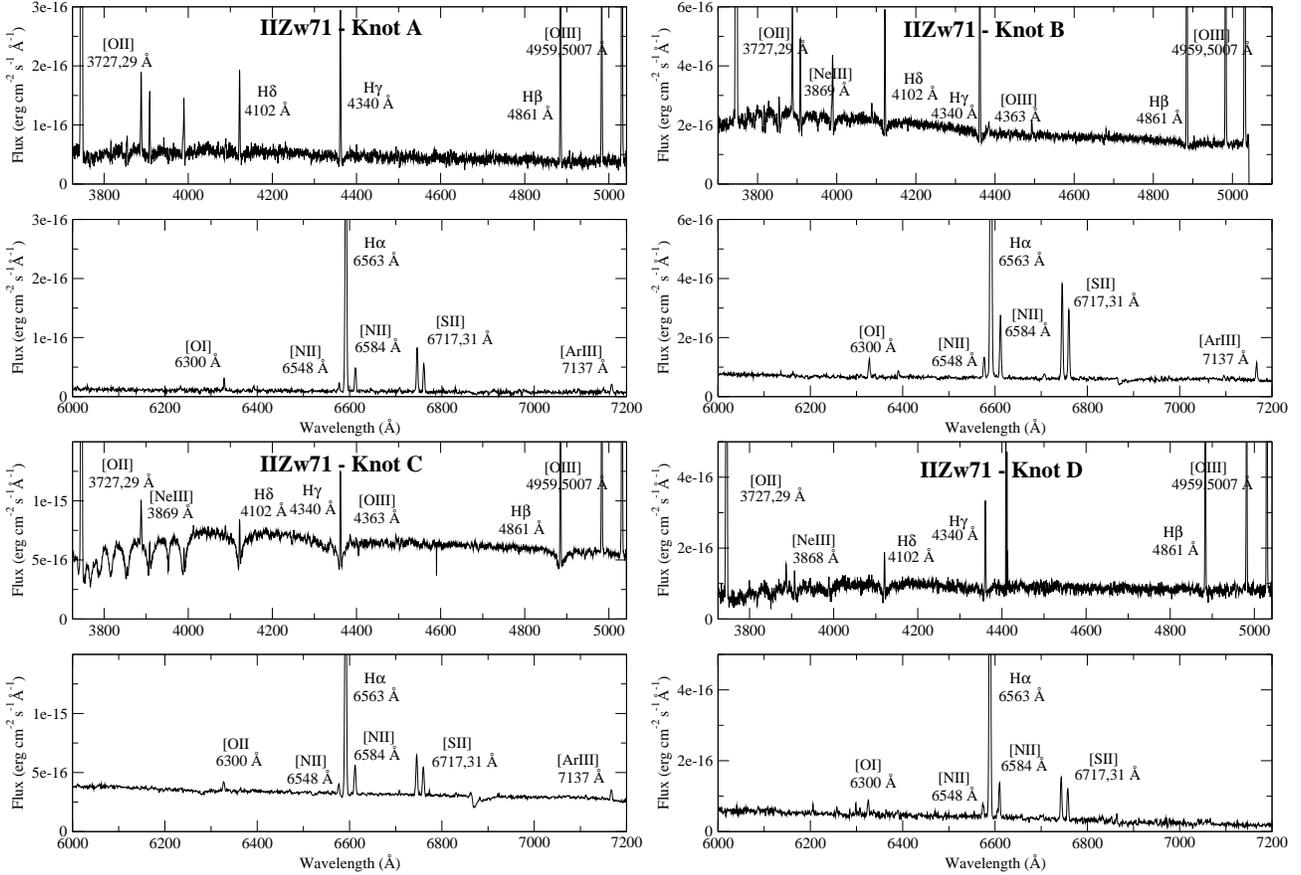

   \centering
   \includegraphics[width=8.5cm,clip=]{0984fg2a.eps}
    \includegraphics[width=8.5cm,clip=]{0984fg2b.eps}
 \includegraphics[width=8.5cm,clip=]{0984fg2c.eps}
 \includegraphics[width=8.5cm,clip=]{0984fg2d.eps}
   \caption{Blue and red spectra up to 7200 {\AA} for each of the extracted knots of IIZw71.}
              \label{spectra}
    \end{figure*}

Figure \ref{spectra} shows the blue and red parts of the spectra of the four observed knots (labelled from A to D), 
which have been extracted from the slit. Knot C, as expected, shows the greatest contribution from an underlying 
stellar population. 

\section{Results and discussion}

\subsection{Line intensities and reddening correction}

The presence of a conspicuous underlying stellar population depresses the Balmer emission lines and does not allow the measurement of their 
fluxes with an acceptable accuracy (D\'\i az, 1988).
To account for this, we subtracted from the observed spectra the best-fitting found by the spectral synthesis code STARLIGHT
\footnote{The STARLIGHT project is supported by the Brazilian agencies CNPq, CAPES and FAPESP and by the France-Brazil CAPES/Cofecub program} (Cid Fernandes et al. 2004, 2005; Mateus
et al. 2005). STARLIGHT fits an observed continuum spectral energy distribution using a combination of multiple simple stellar population (SSPs; also known as instantaneous burst) synthetic spectra using a $\chi^2$ minimisation procedure.
We chose for our analysis the SSP spectra from Bruzual \& Charlot (2003), based on
the STELIB library of Le Borgne et al. (2003), Padova (1994) evolutionary tracks, and a Chabrier (2003)
initial mass function between 0.1 and 100 M$_\odot$. We fixed the metallicity to Z = 0.004 (= 1/5 Z$_\odot$) for all models, which is the metallicity in the Bruzual \& Charlot (2003) library closest to those corresponding to the oxygen abundances measured in knots B and C, as explained in Section 3.3. The code simultaneously finds the ages and the relative contributions of the different present SSPs and the average reddening. The reddening law from Cardelli, Clayton \& Mathis (1989)  was used. Prior to the fitting procedure, the spectra were shifted to the rest frame, and sampled again to a resolution of 1 {\AA}. Bad pixels and emission lines were excluded from the final fits. The percentage of
the mean deviation over all fitted pixels between observed and model spectra ranges between 7.3\% in
knots A and D, which have the lowest signal, to 2.9 \% in knot B and 1.9\% in knot C.

\begin{table}
\centering
\caption[]{Equivalent widths of the hydrogen recombination lines H$\beta$, H$\gamma$ and
H$\delta$ as measured directly in the spectra of the observed knots 
once the underlying stellar populations have been removed, with the corresponding correction factors.}
\label{ews}
\begin{tabular} {c c c c c}
\hline
\hline
Knot  & A & B & C & D         \\
\hline
EW(H$\alpha$) (\AA) & 270$\pm$30 &  130$\pm$10 & 40$\pm$2 & 140$\pm$5 \\
f$_c$   &  1.00 & 1.00 & 1.05 & 1.00 \\
EW(H$\alpha$)$_c$ (\AA) & 270$\pm$30 &  130$\pm$10 & 42$\pm$2 & 140$\pm$5 \\
\hline
EW(H$\beta$) (\AA) & 34.9$\pm$4.5 &  30.0$\pm$1.7 & 10.7$\pm$1.1 & 32.2$\pm$4.7 \\
f$_c$   &  1.01 & 1.03 & 1.13 & 1.00 \\
EW(H$\beta$)$_c$ (\AA) & 35.2$\pm$4.7 &  30.9$\pm$1.8 & 12.1$\pm$1.3 & 32.2$\pm$4.9 \\
\hline
EW(H$\gamma$) (\AA) & 17.2$\pm$1.7 &  12.8$\pm$0.9 & 5.0$\pm$0.9 & 10.8$\pm$1.7 \\
f$_c$   &  1.03 & 1.03 & 1.19 & 1.00 \\
EW(H$\gamma$)$_c$ (\AA) & 17.7$\pm$1.3 &  13.2$\pm$1.0 & 6.0$\pm$1.0 & 10.8$\pm$0.9 \\
\hline
EW(H$\delta$) (\AA) & 8.1$\pm$0.6 &  6.2$\pm$0.7 & 1.8$\pm$1.0 & 8.0$\pm$0.8 \\
f$_c$   &  1.09 & 1.13 & 1.24 & 1.03 \\
EW(H$\delta$)$_c$ (\AA) & 8.8$\pm$0.6 &  7.0$\pm$1.0 & 2.3$\pm$1.0 & 8.2$\pm$1.0 \\
\hline

\hline
\end{tabular}
\end{table}

\begin{figure}[t]
\centering
   \includegraphics[width=8.5cm,clip=]{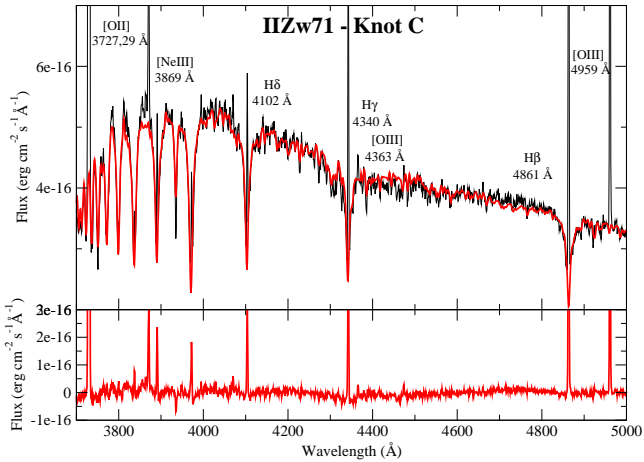}
   \caption{Results of the STARLIGHT fit to the continuum spectral energy distribution for knot C, the one most affected by underlying stellar population contribution. The lower panel shows the emission line spectrum once this contribution has been removed.}
              \label{subtraction}
    \end{figure}

Figure \ref{subtraction} illustrates the results of the model fitting to the spectrum of knot C.
The upper panel shows the observed spectrum and the model spectrum.
The lower panel shows the subtraction of the fitted stellar continuum.
We measured the Balmer emission line fluxes over a linear continuum derived by hand from
the observed spectra. We then compared these with the same measurements over the continuum derived
from the model SSPs. This allows 
the equivalent widths to be corrected by the presence of absorption stellar
features in these lines that cause the measured EWs to be underestimated.
In Table \ref{ews} we summarise our results for the four brightest Balmer emission lines.
For each Balmer line, the first row gives the measured equivalent width of the emission line, while the third row gives the corresponding value after the removal of the underlying absorption. The second row gives the ratio between the two. Knot C is the most affected indicating a larger contribution by the underlying stellar population, although the correction for the H$\beta$ line only amounts to 13\%, reaching 24\% for the H$\delta$ line. For the rest of the knots, only B presents a significant contribution of 13 \% for the H$\delta$ line.


\begin{table*}
\begin{minipage}{170mm}
\caption{Relative observed and reddening corrected line intensities
  [F(H$\beta)$=I(H$\beta)$=1000] for the different observed knots of IIZw71. For each knot we also give the flux of H$\beta$, once reddening corrected and after subtraction of the underlying absorption and the constant of reddening, with their corresponding errors.} 
\label{lineas}
\begin{center}
\begin{tabular}{lcccccccc}
\hline

\multicolumn{1}{c}{$\lambda$  ({\AA})}  & \multicolumn{2}{c}{A} & \multicolumn{2}{c}{B}  & \multicolumn{2}{c}{C} & \multicolumn{2}{c}{D} \\
 & $F(\lambda)$  &  $I(\lambda)$ &  $F(\lambda)$  & $I(\lambda)$ &  $F(\lambda)$  &  $I(\lambda)$ &  $F(\lambda)$  & $I(\lambda)$ \\
\hline

3726  [O{\sc ii}]\footnote{Partially blended (see text and Figure \ref{blend})}        & 2018$\pm$62  &  2070$\pm$240 & 1545$\pm$43 & 1667$\pm$132 &  1497$\pm$60 & 1561$\pm$133 &  1225$\pm$107 &  1552$\pm$227  \\
3729  [O{\sc ii}]$^a$        & 2936$\pm$289 &  2767$\pm$377 & 2271$\pm$61 &  2450$\pm$163 &  1990$\pm$74 & 2076$\pm$156 &  1758$\pm$150 &  2227$\pm$290 \\
3869  [Ne{\sc iii}]        & 321$\pm$37  &  328$\pm$93 & 242$\pm$23 &  258$\pm$50 &  297$\pm$23 & 307$\pm$53 &  183$\pm$25 &  224$\pm$68  \\
4102  H$\delta$              &  303$\pm$29  &  307$\pm$75 &  242$\pm$15  &  254$\pm$40 &   257$\pm$19  & 263$\pm$42 &   233$\pm$32  &  270$\pm$67  \\
4340  H$\gamma$              &  584$\pm$91  &  570$\pm$117 &  466$\pm$13  &  480$\pm$43 &   483$\pm$31  & 491$\pm$52 &   417$\pm$43  &  458$\pm$75  \\
4363  [O{\sc iii}]           & --- & --- & 29$\pm$5 &  30$\pm$11 &  43$\pm$10 & 44$\pm$16 & --- & ---  \\
4861  H$\beta$              &  1000$\pm$39  &  1000$\pm$39 &  1000$\pm$37  &  1000$\pm$37 &   1000$\pm$47  & 1000$\pm$47 &   1000$\pm$118  &  1000$\pm$118  \\
4959  [O{\sc iii}]           & 734$\pm$42 & 733$\pm$54 & 818$\pm$27 &  814$\pm$34 &  937$\pm$34 & 934$\pm$41 &  850$\pm$75 & 837$\pm$80  \\
5007  [O{\sc iii}]           & 2101$\pm$69 & 2096$\pm$99 & 2411$\pm$67 & 2393$\pm$79 & 2770$\pm$95 & 2759$\pm$106 & 2608$\pm$119 & 2550$\pm$223  \\
6300  [O{\sc i}]           &      83$\pm$19    & 81$\pm$30  &  92$\pm$7   & 87$\pm$16 &  77$\pm$6   &  74$\pm$15  &  100$\pm$13  &  84$\pm$25  \\
6548  [N{\sc ii}]           &      101$\pm$19    &  99$\pm$19  &  92$\pm$7    &  86$\pm$8  &   93$\pm$13   & 90$\pm$17  &  100$\pm$13  & 82$\pm$15  \\
6563  H$\alpha$              & 2752$\pm$126 & 2693$\pm$269 & 2895$\pm$93 & 2713$\pm$161 &  3000$\pm$120  & 2894$\pm$183 & 3333$\pm$324 & 2724$\pm$320  \\
6584  [N{\sc ii}]            &  303$\pm$21 &  296$\pm$24 &  250$\pm$18  & 234$\pm$21 &   230$\pm$14  & 222$\pm$18 &  250$\pm$15  & 204$\pm$19  \\
6717  [S{\sc ii}]            &  514$\pm$41  &  502$\pm$59 &  405$\pm$17  & 378$\pm$29 &   297$\pm$12  & 285$\pm$24 &   267$\pm$15  &  215$\pm$29  \\
6731  [S{\sc ii}]            &  339$\pm$30  & 332$\pm$47 & 292$\pm$12  & 272$\pm$24 &   217$\pm$14  & 208$\pm$24 &  200$\pm$9  & 161$\pm$24  \\
7137  [Ar{\sc iii}]            &  138$\pm$19  & 134$\pm$46 &  68$\pm$7  & 63$\pm$18 &   77$\pm$6  & 73$\pm$20 & ---  &  ---  \\
9069  [S{\sc iii}]           &  92$\pm$28 & 88$\pm$42 &  124$\pm$15  & 110$\pm$43 &  107$\pm$14   & 100$\pm$43  &  250$\pm$43   &  174$\pm$80  \\
\hline
\multicolumn{1}{c}{F(H$\beta$)(10$^{-15}$erg\,seg$^{-1}$\,cm$^{-2}$) }  & \multicolumn{2}{c}{0.89$\pm$0.05} & \multicolumn{2}{c}{3.61$\pm$0.10} & \multicolumn{2}{c}{4.19$\pm$0.10} & \multicolumn{2}{c}{3.49$\pm$0.10} \\
\multicolumn{1}{c}{c(H$\beta$)}  & \multicolumn{2}{c}{$<$0.05} & \multicolumn{2}{c}{0.09$\pm$0.03} & \multicolumn{2}{c}{0.05$\pm$0.02} & \multicolumn{2}{c}{0.28$\pm$0.02} \\

\hline

\end{tabular}
\end{center}
\end{minipage}
\end{table*}

The  emission line fluxes were measured on the spectra using the \texttt{splot} task of IRAF and are
listed for the four observed knots in Table \ref{lineas}. 
Column 1 of the table shows the wavelength and identification of the
measured lines. The observed
emission line fluxes, $F(\lambda)$ (in  units of  H$\beta$ flux = 1000)
with their corresponding errors, are given in columns 2, 4, 6, and 8 for Knots A, B, C, and D respectively.
Only the fluxes of the Balmer recombination lines were measured on the subtracted spectra, while
the collisional emission lines were measured on the uncorrected spectra in order to reduce the
uncertainties due to the subtraction of the continuum.

We used two different ways to integrate the flux of
a given line. (1) In the case of an isolated line, the intensity was calculated by integrating between two points given by the position of a local continuum placed by eye. (2) In the case of the two lines of [O{\sc ii}] at $\lambda\lambda$ 3727, 29 \AA, which are partially blended,  a multiple Gaussian fit procedure to estimate individual fluxes was used. This procedure is illustrated in Figure \ref{blend}. The statistical errors associated with the measured emission line fluxes were calculated using the expression
\[ \sigma_{l}\,=\,\sigma_{c}N^{1/2}[1 + EW/(N\Delta)]^{1/2} \]
\noindent where $\sigma_{l}$ is  the error in the measured line flux, $\sigma_{c}$ represents the standard deviation in a box near the measured emission line and stands for the error in
the continuum placement, $N$ is the number of pixels used in the measurement of 
the line flux, $EW$ is the line equivalent width, and $\Delta$ is the wavelength
dispersion in \AA ngstroms per pixel.



The reddening coefficients ($c$(H$\beta$) were calculated from the measured Balmer decrements, $F(\lambda)$/$F(H\beta)$, adopting the galactic extinction law of Cardelli et al. (1989) with $R_{\rm v}$=3.1. A least square fit of the measured decrements to the theoretical ones,  $(F(\lambda)$/$F(H\beta))_0$,
computed based on the data by Storey \& Hummer (1995), was performed that provides the value of $c$(H$\beta$). The theoretical Balmer decrements depend on electron temperature
and density. We used an iterative method to estimate them, taking as starting values those derived from the measured [S{\sc ii}] $\lambda\lambda$ 6717,6731 {\AA}  and  [O{\sc iii}]  $\lambda\lambda$ 4363, 4959, 5007 {\AA} line fluxes.  

\begin{figure}[t]
   \centering
   \includegraphics[width=8.5cm,clip=]{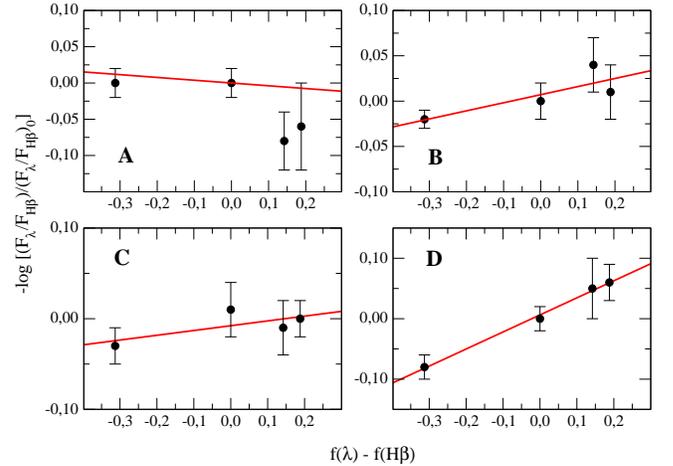}
 
   \caption{Representation for each observed knot of the ratio of the expected
and observed ratios of the Balmer hydrogen lines and the extinction curve. The slope of
the obtained fit is the constant of reddening.}
              \label{red}
    \end{figure}

\begin{figure}[t]
\centering
   \includegraphics[width=8.5cm,clip=]{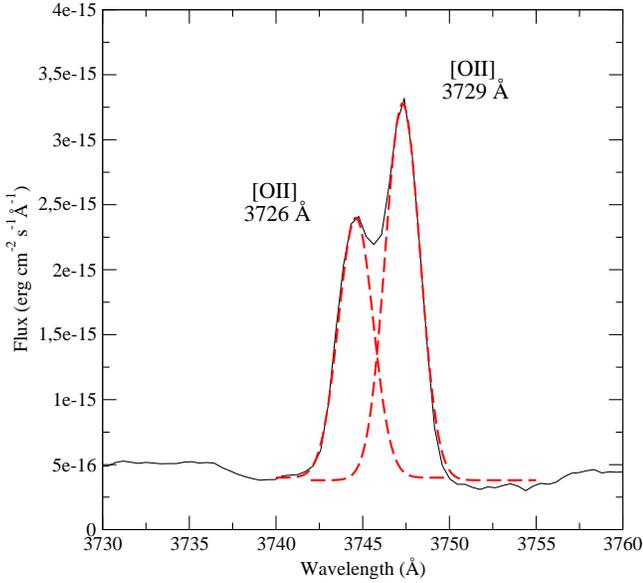}
   \caption{Deblending of the [O{\sc ii}] $\lambda\lambda$ 3727, 3729 \AA\ lines. Their ratio allows to put lower limits to the electron density in the observed ionized regions.}
              \label{blend}
    \end{figure}

In Figure \ref{red} we show the actual fits for each of the observed knots. The 
Galactic extinction for this object is negligible, so we can be sure that 
internal reddening is the unique source of extinction.  All the observed knots
present positive values of c(H$\beta$), except knot A in which the extinction is consistent with zero value. The highest value of c(H$\beta$) is found for knot D.

The values of c(H$\beta)$) and their corresponding errors, considered as
the uncertainties of the least square fits, are listed at the bottom of Table
\ref{lineas} for the observed knots, together with the reddening corrected H$\beta$ intensity.  The
corrected intensities for the observed lines relative to H$\beta$, $I(\lambda)$, are given in columns 3, 5, 7, 
and 9 of the same table. In all cases, the reddening correction was done relative to the closest Balmer
recombination line.

\begin{figure}[t]
\centering
   \includegraphics[width=8.5cm,clip=]{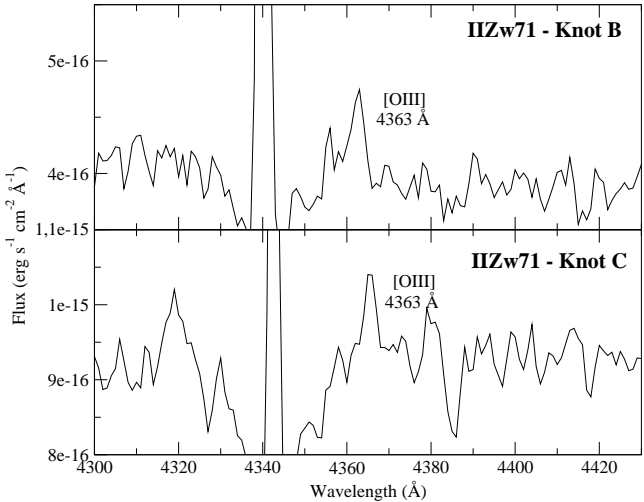}
   \caption{Detail of the rest-frame spectra of the knots B and C around the [O{\sc iii}] 4363 {\AA} emission line.}
              \label{4363}
    \end{figure}

\subsection{Physical conditions of the gas}

The physical conditions of the ionized gas, including electron temperatures and
electron densities, were derived from the emission line data using the same
procedures as in P\'erez-Montero \& D\'\i az (2003), based on the five-level
statistical equilibrium atom approximation in the task \texttt{temden} of the software package IRAF
(De Robertis, Dufour \& Hunt, 1987; Shaw \& Dufour, 1995). The atomic
coefficients used here are the  same as in H\"agele et al. (2006). We took as sources of error the
uncertainties associated with the measurement of the emission-line fluxes and
the reddening correction and we propagated them through our calculations. 
Electron densities were derived in the four knots from both 
[O{\sc ii}]\,$\lambda\lambda$\,3726\,/\,3729\,\AA\ and
[S{\sc ii}]\,$\lambda\lambda$\,6717\,/\,6731\,\AA\ line ratios, which are
representative of the low-excitation zone of the ionized gas. The values
for each of the knots are listed in Table \ref{tab_dens}. In all cases they provide upper limits, which are lower when the [O{\sc ii}] lines are used. In the case of these lines, however, the spectral dispersion of our data do not allow total resolution of the lines, and they were deconvolved by fitting two Gaussian components as shown in Figure \ref{blend}. The upper limits for the electron density are lower in all cases than the critical value.


\begin{table}
\centering
\caption[]{Electron densities in particles per cm$^3$ of [S{\sc ii}] and [O{\sc ii}] measured for each of the knots.}
\label{tab_dens}
\begin{tabular} {l c c}
\hline
 Knot & n([O{\sc ii}])  & n([S{\sc ii}]) \\
 \hline
A & $<$200 & $<$230 \\
B & $<$85 & $<$160 \\
C & $<$180 & $<$210 \\
D & $<$190 & $<$350 \\
\hline
\end{tabular}
\end{table}

For knots B and C, for which the intensity of the [O{\sc iii}]$\lambda$ 4363 {\AA}
was measured with sufficient signal-to-noise (see Figure \ref{4363} for a more detailed plot
of this line in these knots), the [O{\sc iii}] electron temperature was derived directly from the ratio
 (I(4959\AA)+I(5007\AA))/I(4363\AA). Then we derived T([O{\sc ii}]) from T([O{\sc iii}]) using the relations based on the photoionization models described in P\'erez-Montero \& D\'\i az (2003), which take explicitly nto account 
the dependence of T([O{\sc ii}]) on the electron density, and T([{\sc iii}]) has been estimated from the empirical relation:

\[T([S\textrm{\sc iii}]) = (1.19 \pm 0.08)\,T([O\textrm{\sc iii}]) - (0.32 \pm 0.10)\]

\noindent found by H\"agele et al. (2006) for HII galaxies.  The values of the derived electron 
temperatures for these two knots are listed in Table \ref{ionic}.

\begin{table}
\begin{minipage}{85mm}
{\small
\caption{Electron temperatures, ionic and total abundances for
the knots B and C of IIZw71.} 
\begin{center}
\begin{tabular}{l cc }
\hline
\hline
Knot & B & C \\
                          
\hline                                    
T([O{\sc iii}])&  12400$\pm$1700 K & 13700$\pm$2000 K     \\ 
T([O{\sc ii}])\footnote{From a relation with T([O{\sc iii}]) based on photoionization models} &  12900$\pm$1300 K & 12800$\pm$1300 K       \\ 
T([S{\sc iii}])\footnote{From an empirical relation with T([O{\sc iii}])} &  11600$\pm$1100 K & 13200$\pm$1200 K     \\ 
\hline
12+$\log(O^+/H^+)$    &7.67$\pm$0.23 & 7.63$\pm$0.23   \\
12+$\log(O^{2+}/H^+)$ & 7.63$\pm$0.19 & 7.57$\pm$0.19  \\
\bf{ 12+log(O/H)}     & 7.95$\pm$0.21 & 7.90$\pm$0.21 \\
\hline                                                                  
12+$\log(S^+/H^+)$    & 5.94$\pm$0.12 & 5.83$\pm$0.13  \\
12+$\log(S^{2+}/H^+)$ & 6.00$\pm$0.26 &  5.85$\pm$0.26  \\
ICF($S^++S^{2+}$)     & 1.15$\pm$0.13 & 1.14$\pm$0.12 \\
\bf{ 12+log(S/H)}     & 6.33$\pm$0.24 & 6.20$\pm$0.24   \\
\hline                                                                  
12+$\log(N^+/H^+)$    & 5.87$\pm$0.28 & 5.63$\pm$0.23   \\
\bf{log(N/O)}         & -1.48 $\pm$0.37 & -1.68 $\pm$0.32   \\
\hline                                                                  
12+$\log(Ne^{2+}/H^+)$& 7.12$\pm$0.29 & 7.06$\pm$0.28   \\
ICF($Ne^{2+}$)& 1.22$\pm$0.13 & 1.23$\pm$0.14   \\
\bf{ 12+log(Ne/H)}    & 7.19$\pm$0.35 &  7.13$\pm$0.35   \\
\hline                                                                  
12+$\log(Ar^{2+}/H^+)$& 5.61$\pm$0.22 &  5.57$\pm$0.19  \\
ICF($Ar^{2+}$)& 1.16$\pm$0.03 & 1.16$\pm$0.04   \\
\bf{ 12+log(Ar/H)}    & 5.65$\pm$0.26 &  5.60$\pm$0.26   \\
\hline                                                        
\hline

\end{tabular}
\label{ionic}
\end{center}}
\end{minipage}
\end{table}


\subsection{Ionic and total chemical abundances in knots B and C.}

The oxygen ionic abundance ratios, O$^{+}$/H$^{+}$ and O$^{2+}$/H$^{+}$, 
were derived from the [O{\sc ii}]\,$\lambda\lambda$\,3727,3729\,\AA\ and [O{\sc
    iii}] $\lambda\lambda$\,4959, 5007\,\AA\ lines, respectively using the
appropriate electron temperatures for each ion. The total
abundance of oxygen was derived assuming
\[ 
\frac{O}{H}\,\approx\,\frac{O^++O^{2+}}{H^+}
\]

We derived S$^+$ abundances from  the fluxes of the [S{\sc ii}] emission lines at $\lambda\lambda$\,6717, 6731\,{\AA} assuming  that T([S{\sc ii}]) $\approx$ T([O{\sc ii}]) and S$^{2+}$ abundances have been derived from the fluxes of the near-IR [S{\sc iii}]\,$\lambda$\,9069 line and the estimated value of T([S{\sc iii}]).
The total sulphur abundance was calculated using an ionization correction factor (ICF) for S$^+$+S$^{2+}$ according to Barker's (1980) formula, which is based on the photoionization
models of Stasi\'nska (1978): 
 \[
ICF(S^++S^{2+}) \approx \left[ 1-\left( \frac{O^{2+}}{O^++O^{2+}}
  \right)^\alpha\right]^{-1/\alpha} 
\]
\noindent where $\alpha$\,=\,2.5 provides the best fit to the scarce observational
data on S$^{3+}$ abundances (P\'erez-Montero et al.\ 2006).  Taking this ICF as
a function of the ratio O$^{2+}$/O instead of O$^+$/O reduces the propagated error
for this quantity. The ICF found are similar for the two knots and increase the derived S$^+$+S$^{2+}$ by about 15 \% .
 
The ionic abundance of nitrogen, N$^{+}$/H$^{+}$, was derived from the
intensities of the $\lambda\lambda$\,6548, 6584\,\AA\ lines assuming that  T([N{\sc ii}]) $\approx$ T([O{\sc ii}]).
Then, the total N abundance was derived under the following
assumption:
\[
\frac{N}{O}\,\approx\,\frac{N^+}{O^+}
\]

Neon is only visible in the spectra by means of the [Ne{\sc iii}] emission line at
$\lambda$3869\,{\AA}. For this ion we took the electron temperature 
of [O{\sc iii}], as
representative of the high excitation zone. The total abundance of neon was 
calculated using the following expression for the  ICF (P\'erez-Montero et al., 2007):
\[
ICF(Ne^{2+}) \approx 0.753 + 0.142\cdot  \frac{O^{2+}}{O^++O^{2+}} + 0.171 \cdot \frac{O^++O^{2+}}{O^{2+}}
\]
This formula considers the overestimate of
Ne/H in objects with low excitation, where the charge transfer between O$^{2+}$
and H$^0$ becomes important (Izotov et al., 2004). 
 
The only accesible emission lines of argon in the optical spectra of ionized regions correspond to  Ar$^{2+}$ and
Ar$^{3+}$. In the spectra of knot B and C, only [Ar{\sc iii}]\,$\lambda$\,7136\,\AA\ was measured and the abundance of Ar$^{2+}$ was 
calculated assuming that T([Ar{\sc iii}])\,$\approx$\,T([S{\sc iii}]) 
(Garnett, 1992). The total abundance of Ar was then calculated  using the 
ICF(Ar$^{2+}$) derived from photo-ionization models by P\'erez-Montero et al. (2007): 
\[
ICF(Ar^{2+})\,=0.749+0.507\cdot\Big(1-\frac{O^{2+}}{O^++O^{2+}}\Big)+\]
\[
+0.0604\cdot\Big(1-\frac{O^{2+}}{O^++O^{2+}}\Big)^{-1}
\] 
\noindent The calculated ICF for Ar and Ne are the same in the two knots within the errors.
The ionic and total abundances for the observed species in knots B and C are given in
Table \ref{ionic}, along with their corresponding errors.   

\subsection{Metal content of the polar ring}


   \begin{figure*}
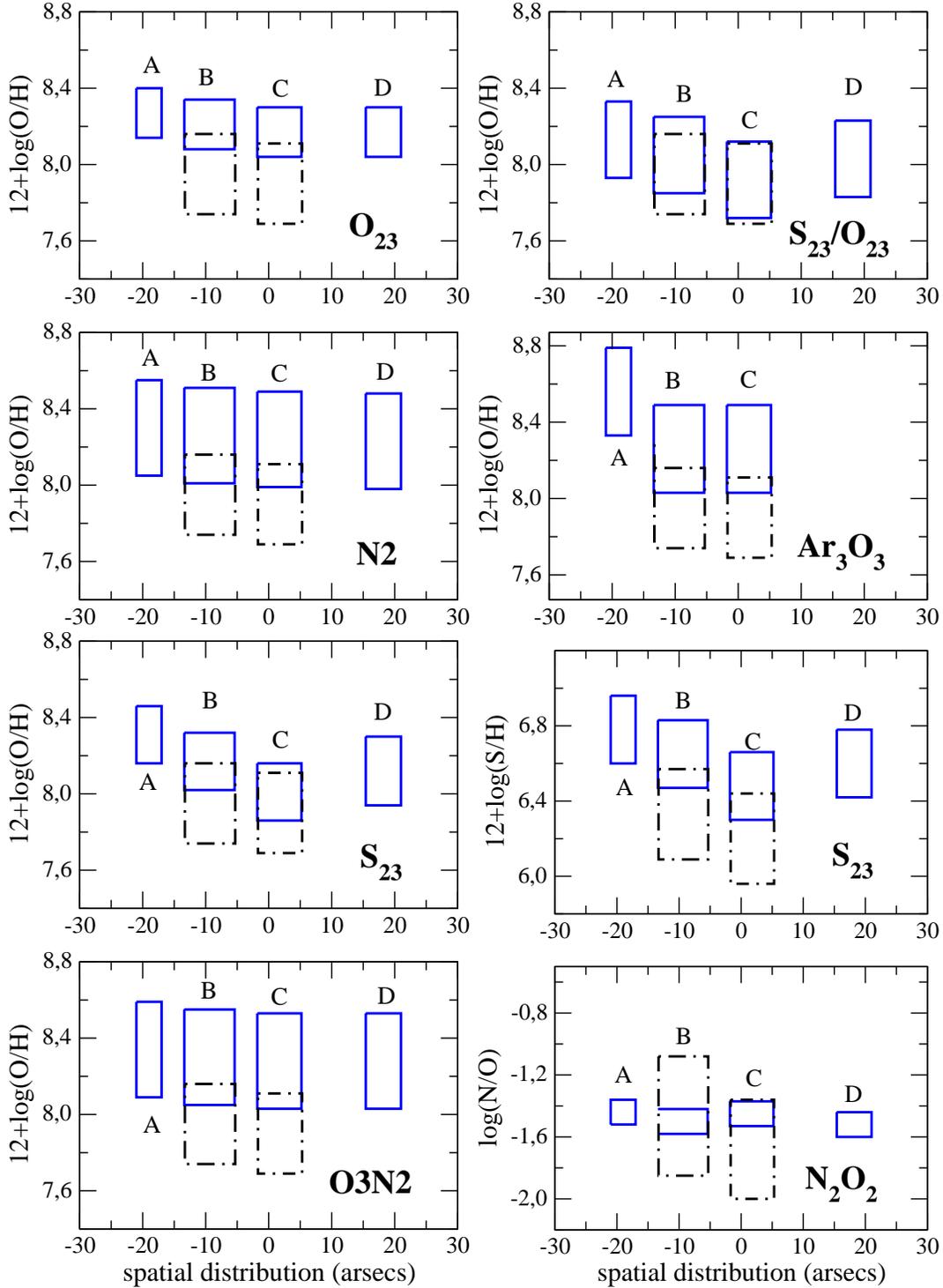

   \centering
  \includegraphics[width=7cm,clip=]{0984fg7a.eps}
  \includegraphics[width=7cm,clip=]{0984fg7b.eps}
\includegraphics[width=7cm,clip=]{0984fg7c.eps}
  \includegraphics[width=7cm,clip=]{0984fg7d.eps}
\caption{The oxygen abundances and their uncertainties for each observed knot, as derived
using different empirical calibrators. From upper to lower panel: O$_{23}$, N2, S$_{23}$ and O3N2, at
left and Ar$_3$O$_3$ and S$_{23}$/O$_{23}$ at right. The lower panels of right column represent the sulphur abundance
as derived from the S$_{23}$ parameter and the N/O ratio as derived from the N$_2$O$_2$ parameter. The dash-dotted
line represent the abundances and their uncertainties as derived from the direct method in knots B and C.}
\label{met}
 \end{figure*}

The emission-line spectra of the four star-forming knots in IIZw71 are remarkably similar, implying similar values for ionization parameter, ionization temperature, and chemical abundances. We derived the ionization parameters from the ratio of the [O{\sc ii}] and [O{\sc iii}] lines according to the expression given in D\'\i az et al. (2000). They are similar in all the knots ranging from 6.42$\times$10$^{-4}$ for knot A to 9.67$\times$10$^{-4}$ for knot C. Using these values, the corrected H$\alpha$ fluxes, and the sizes of the regions from H$\alpha$ images we can calculate the density of the emitting gas (D\'\i az et al. 1991). This is similar for the four knots with a value of about 20 particles $\cdot$ cm$^{-3}$, consistent with the upper limits we derived from the ratio of the [S{\sc ii}] lines and providing filling factor for the gas of a few times 10$^{-2}$, values common to giant HII regions.

We derived oxygen, sulphur, nitrogen, neon, and argon chemical abundances in knots B and C, where
the auroral line of [O{\sc iii}] at 4363 {\AA} had been measured and thus the electron temperature of the high-excitation zone
also was derived. We assumed also an inner structure of the electron
temperature to deduce the values of the temperatures of [O{\sc ii}], which is representative of the low-excitation
zone, affecting the calculation of the chemical abundances of O$^+$, N$^+$, and S$^+$. This inner ionization
structure leads to derivation of the temperature of [S{\sc iii}], 
valid for the intermediate zone (Garnett, 1992) and calculation of the ions of S$^{2+}$ and Ar$^{2+}$. 
The derived ionic abundances in these two knots
are very similar, leading to total O/H abundances of 12+log(O/H) = 7.90 in knot C and 7.95 in knot B.
These values deviate slightly from the metallicity reported in the literature for
the integrated object, which is 12+log(O/H) = 8.24 (Shi et al., 2005; Kewley et al., 2005) and the difference is even greater than 
the value for the nuclear region, which could be considered as knot C, given by Kewley et al. (2005), which is 12+log(O/H) = 8.55. The value
reported by Shi et al. (2005) are based on the direct method, while those reported by Kewley et al. (2005) were derived using empirical calibrations since no temperature sensitive lines were observed.

This is also the case in our observations of knots A and D. The calculation of metallicities in these two knots can only be carried out with  calibrations based on
the strongest emission lines because no auroral lines were detected and the direct method cannot be used. 

The different strong-line methods for abundance derivations, which have been
widely studied in the literature, are based on the directly calibrating of the relative
intensity of some bright emission lines against the abundance of some relevant ions
present in the nebula. For the case of oxygen, 
we take the calibrations studied by
P\'erez-Montero \& D\'\i az (2005), who obtain different uncertainties for each parameter
in a sample of ionized gaseous nebulae with accurate determinations of chemical abundances
in the whole range of metallicity. 
 
In Figure \ref{met}, we show the corresponding total abundances as derived from several strong-line methods and the oxygen abundances calculated from the electron
temperatures measured in knots B and C.
Among the available strong-line parameters we studied the O$_{23}$ parameter (also known
as R$_{23}$ and originally defined by Pagel et al. (1979) and based on [O{\sc ii}] and [O{\sc iii}] strong emission lines). This parameter is characterised by
its double-valued relation with metallicity, with a very large dispersion in the turnover region. According to the values measured in knots B and C, we used the McGaugh (1991) calibration for the lower branch, obtaining similar values for the oxygen abundance in the four observed knots.

The N2 parameter (defined by Storchi-Bergmann et al., 1994) is based on the strong emission lines of [N{\sc ii}]. It remains single-valued up to high metallicities in its relation to oxygen abundance, and it is almost independent of reddening and flux calibrations. Nevertheless, it has the high dispersion associated to the functional parameters of the nebula (ionization parameter and ionizing radiation temperature) and to N/O variations. We used the empirical calibration of this parameter from Denicol\'o et al. (2002) to derive the oxygen abundance in the four knots of this galaxy. We can see in Figure \ref{met} that the abundances predicted by this parameter are quite similar for all the knots although more than the values derived from the direct method in knots B and C.

The S$_{23}$ parameter was defined by V\'\i lchez \& Esteban (1996) and is based on the strong
emission lines of [S{\sc ii}] and [S{\sc iii}]. The calibration done by P\'erez-Montero \& D\'\i az (2005) yields oxygen abundances comparable to the four observed knots, slightly higher than the directly derived abundances of knots B and C, but still consistent with them within the errors.  In the case of sulphur, both the directly and empirically derived abundances of knots B and C are slightly different with knot C showing a lower abundance. However it should be remembered that the near IR sulphur lines relative to the hydrogen recombination lines are more affected by reddening than in the case of oxygen when no Paschen lines are observed, which is the case. 

The parameter O3N2, defined by Alloin et al. (1979) depends on strong emission lines of [O{\sc iii}] and [N{\sc ii}]. We used the calibration due to Pettini \& Pagel (2004) and, as we can see in Figure \ref{met} it has a very similar behaviour to that of N2. 

The combination of the S$_{23}$ and O$_{23}$ parameters gives S$_{23}$/O$_{23}$, defined by D\'\i az \& P\'erez-Montero (2000), a parameter that increases monotonically with the oxygen abundance up to oversolar regime and which is very useful to study variations over wide ranges of metallicity ({\em e.g.} disks). We applied the calibration found in P\'erez-Montero \& D\'\i az (2005) and, in this case, there is almost a perfect coincidence with the values found by the direct method in knots B and C. Again, we find a similar value in knots D and A.

The Ar$_3$O$_3$ parameter, defined and calibrated by Stasi\'nska (2006) as the ratio of [Ar{\sc iii}] 7136 {\AA} and [O{\sc iii}] 5007 {\AA} emission lines, predicts identical values for the metallicity in knots B and C, although noticeably higher than the directly derived ones. A rather high value is found in knot A and no data on the [Ar{\sc iii}] line in knot D were available.

Finally, the N$_2$O$_2$ parameter, defined by P\'erez-Montero \& D\'\i az  (2005) as the ratio between [N{\sc ii}] and [O{\sc ii}] emission lines, can be used to obtain the N/O ratio. Using this parameter, very little difference is found in N/O among the four observed knots, which are also consistent with the directly derived ratio.

\begin{figure*}[t]
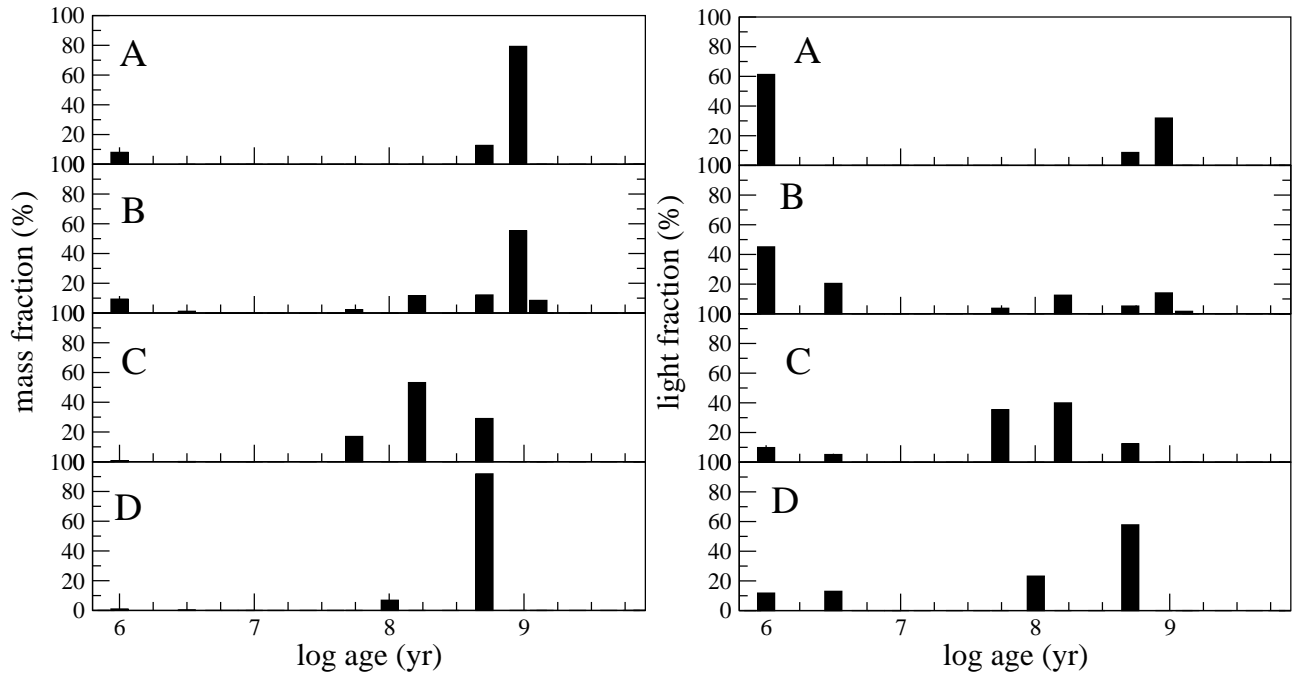

\centering

\includegraphics[width=8.5cm,clip=]{0984fg8a.eps}
\includegraphics[width=8.5cm,clip=]{0984fg8b.eps}
\caption{Histogram of the distribution in mass fraction (left) and visual light (right) of the most probable stellar population models fitted by STARLIGHT for each of the observed knots.}
\label{starlight}
\end{figure*}


\begin{table}
\caption{Values of the extinction, total stellar mass, and fraction of the mass in stars younger than 10 Myr for each knot in the best-fit model predicted by STARLIGHT.}
\centering
\begin{tabular}{l cc c }
\hline
\hline
ID           &   c(H$\beta$) & log (M/M$_\odot$)  & f(age $<$ 10 Myr) \\ 
 \hline
A            &  0.19 & 6.13 $\pm$ 0.05 & 0.08 \\ 
B            &  0.46 & 6.92 $\pm$ 0.05 & 0.11 \\  
C            &  0.04 & 7.52 $\pm$ 0.04  &  0.008  \\ 
D            &  0.25 & 6.46 $\pm$ 0.04 & 0.013 \\          
\hline
\end{tabular}
\label{mass}
\end{table}

\subsection{The stellar population}

The study of the stellar content in the four observed knots was carried out using
STARLIGHT code, which calculates
the combination of stellar libraries and the extinction law that reproduce the spectral distribution 
of energy, to derive the properties of the stellar population in each of the knots better. The used Population Synthesis
stellar libraries and the method to fit the observed spectra of each of the knots
in the polar ring is described in Sect. 3.1.

In Figure \ref{starlight} we show the age distributions of the mass fraction (left) and the visual light fraction 
(right) for the four knots.
All of them present a very young stellar populations with ages below 10 Myr, responsible of the
ionization of the surrounding gas, although most of the stellar mass belongs to populations older than 100 Myr in almost all cases. The estimated total stellar mass and the fraction by mass of the stellar population with an age younger than 10 million years, responsible for the ionization of 
the gas, are listed in Table \ref{mass} for all the knots, together with the internal extinction, represented by the reddening constants, estimated by the model.  The masses were corrected using an aperture factor, which takes
into account the light not collected by the slit and which was calculated as the weighted mean of the ratios of
the light in the whole region as measured in the H$\alpha$ image and in the regions covered by the slit.

Although it is not expected to find the same values of extinction in the gas and the stellar population, there is a very good agreement between the values of the extinction estimated by STARLIGHT in knots C and D in relation to
the values obtained from the decrement of Balmer. In knots A and B, the values are higher than those derived for the ionized gas. In these two knots the specific weight of the youngest population is higher. Since STARLIGHT assumes a common source of extinction for all the stellar populations entering the fit, the degeneracy between age and extinction tends to overestimate the extinction derived for the optimal fit. On the other hand, in knots C and D, which are probably contaminated by the light coming from the host galaxy, the contribution by the young stellar population is much lower, and the light at longer wavelengths is dominated by the older stellar population and the estimated average extinction is more representative.

\begin{center}
\begin{table*}[t]

\caption{Derived properties of the observed knots using the extinction-corrected H$\alpha$ fluxes 
measured in the images from Gil de Paz et al. (2003). }

\centering
\begin{tabular}{c c c c c c c c}
\hline
\hline
ID & R$_{eq}$ & I(H$\alpha$) & L(H$\alpha$) & Q(H$^o$) & M$_{ion}$ & M(HII)  & SFR \\
  & (arcsec)    & 10$^{-14}$erg $\cdot$ s$^{-1} \cdot$ cm$^{-2}$)  &  (10$^{38}$ erg $\cdot$ s$^{-1}$) &  (10$  ^{51}\cdot$s$^{-1}$) & (10$^5\cdot$M$_{\odot}$)  & (10$^4\cdot$M$_{\odot}$)   &  M$_{\odot}$/yr\\
\hline

A  &  1.64 & 1.25  & 4.91 &   3.59 & 7.16 & 3.64 & 0.004 \\
B  &  2.94 & 4.83  & 19.0 & 13.9   & 27.6 & 14.1 & 0.015 \\
C  &  2.65 & 3.96  & 15.6 & 11.4   & 22.7 & 11.5 & 0.012 \\
D  &  1.43 & 1.10  & 4.33 & 3.16   & 6.31 & 3.21 & 0.003 \\      
\hline
\end{tabular}
\label{prop}
\end{table*} 
\end{center}

\subsection{Ionising stellar populations}

Some properties of the emission knots can be obtained from the measured H$\alpha$ flux, such as H$\alpha$ luminosity, number of ionising photons, mass of ionising stars and mass of ionised hydrogen (see D\'\i az et al. 2000). In order to obtain these quantities, we analysed the H$\alpha$ image of the galaxy, retrieved from the
Palomar/Las Campanas atlas of blue compact dwarf galaxies (Gil de Paz et al., 2003). We defined
elliptical apertures on this image for each of the four knots extracted in the spectroscopic observations 
and we measured all the flux inside the elliptical apertures up to the isophote corresponding to the 30\% of
the light in the brightest knot. These regions are shown in the right panel of Figure 1, and their sizes are 
given in the first column of Table \ref{prop}, as the radius of a circular aperture of area equal to that
encompassed by the refereed isophote.
The observed H$\alpha$ flux was corrected in each knot for reddening using the values of the reddening constants,
c(H$\beta$), given in Table \ref{lineas}. The derived values are given in Table \ref{prop}.

The derived values of the masses of the ionising clusters can be compared with those provided by the
STARLIGHT fit. STARLIGHT gives lower values than the ones we derive by factors between 6 and 8 for knots A, B and, C. 
In the case of knot D, the difference is more than an order of magnitude. If we 
take the masses derived from the H$\alpha$ fluxes, which are, in principle, lower limits since neither dust absorption nor a possible escape of photons are taken into account, and we use the calculated proportions of young to total mass given in Table \ref{mass} we obtain total masses between 0.9 $\times$ 10$^7$ and 4.9$\times$ 10$^7$ M$_{\odot}$ for knots A, B and, D and 2.8$\times$10$^8$ M$_{\odot}$ for knot  C. This high value points to a large contribution to the older population by the bulge of the host galaxy. This is consistent with the continuum light distribution shown in Figure \ref{profile} which is wider than that associated with cluster C and peaks at a region displaced from that of the cluster by 1.6 arcsec (about 144 pc). On the other hand for knot B the cluster shows a significant continuum of the same spatial extent as the line emission and the maxima of both distributions coincide. Knots A and D show very little continua of their own, so their underlying population probably corresponds to the outer parts of the host galaxy.

Since the differences in metallicity among the different knots are not significant and the age distributions are also similar, this could indicate a common chemical evolution of these knots, probably related to the process of interaction with the companion galaxy IIZw70. 

The star formation rate (SFR) for each knot was derived from the H$\alpha$ luminosity using the expression given by Kennicutt (1998): 
\[ SFR = 7.9 \times 10^{-42} \times L(H\alpha) \].
The derived values are also given in Table \ref{prop}. They vary from 0.003 M$_\odot$ yr$^{-1}$ for  knot D to 0.015 M$_\odot$ yr$^{-1}$ for knot B and provide a total SFR for the ring of 0.035 M$_ \odot$ yr$^{-1}$, about half of the values quoted by  Kewley et al. (2005)  for the integrated object. which is  0.066 M$_{\odot}$ yr$^{-1}$. Using the sizes of the emitting regions obtained from the H$\alpha$ images and listed in Table \ref{prop}, we obtain a rather constant SFR per unit area, of the order of 7$\times$10$^{-8}$ M$_{\odot}$ yr$^{-1}$ pc$^{-2}$. This is higher than the average value of individual HII regions in polar rings given in Reshetnikov \& Combes (1994) which is 3.2 $\times$10$^{-9}$ M$_{\odot}$ yr$^{-1}$ pc$^{-2}$.

\subsection{Kinematics and dynamics of the polar ring} 

   \begin{figure}
   \centering
  \includegraphics[width=9cm,clip=]{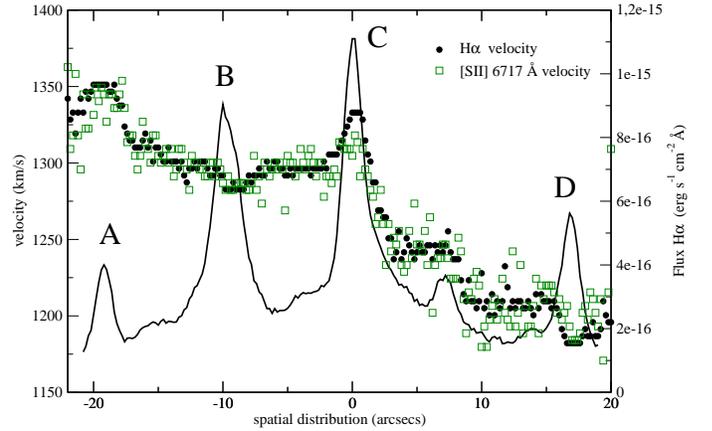}
 
\caption{Rotation curve of IIZw71. Filled circles represent heliocentric velocity derived from H$\alpha$ and open squares from [S{\sc ii}] 6717 {\AA}. Solid line represents the spatial distribution of flux of H$\alpha$ to be compared with the rotation curve.}
\label{kin}
 \end{figure}

We have analysed the differential radial velocity along the slit using the
H$\alpha$ and [S{\sc ii}] 6717 {\AA} emission lines.
The heliocentric velocities are displayed in Figure \ref{kin} superimposed on the observed H$\alpha$ profile.
We find an asymmetric rotation, as it was already stated by Reshetnikov \& Combes (1994), possibly
affected by the expanding velocities of the bubbles of ionised gas surrounding the knots
of star formation which reaches 60 km/s in the case of knot C.

Assumming that we see the polar ring of the galaxy edge-on, with a radial velocity
of 85 km/s, and considering an optical radius of 20 arcsec, which is where we can measure the emission lines with a good enough signal-to-noise ratio, we calculated a dynamical mass inside this
radius of (2.8 $\pm$ 0.2) $\times$10$^9$ M$_{\odot}$. We then calculated the L$_B$ luminosity inside the radius of 20 arcsec, internal to the polar ring, from the B brightness distribution given in Cairos et al. (2001), obtaining L$_B$ = 7.2$\times$10$^8$ L$_{\odot}$. Therefore, we obtain a value of 3.9 for the $M/L_B$ ratio inside the star forming-ring.
This is close to the value of 2.8 found by 
Reshetnikov \& Combes (1994) from optical observations within a distance of 30 arcsec from the centre. 
Considering that their considered blue light probably encompasses the light coming from the emission knots in the ring, that  luminosity  probably constitutes an upper limit and provides a lower limit to the $M/L_B$.

\section{Summary and conclusions}

Spectrophotometric observations of the galaxy IIZw71 were carried
out to study the physical properties of the bursts of star formation along its
polar ring. We extracted information
on the four brightest knots in H$\alpha$ that we labelled from A to D.

The electron temperature of [O{\sc iii}] was measured in knots B and C,
allowing the direct derivation of ionic abundances of oxygen, sulphur, nitrogen, neon and
argon. The total abundances of these species are in the same range of metallicities
measured in HII galaxies, but they are slightly lower than the abundances previously reported
from measurements of the integrated galaxy acording to strong-line calibrations. The metallicity in the other two fainter knots, where the temperature sensitive line could not be detected, was estimated by means of different strong-line parameters. In all cases, the estimated abundances are consistent with those derived for knots B and C by the direct method with the parameter involving the sulphur lines  providing the abundance closer to it. The rest of the parameters slightly overestimate the oxygen abundance. The N/O abundance, as derived from the N$_2$O$_2$ parameter (the ratio of the [N{\sc ii}] and [O{\sc ii}] intensities), is remarkably constant over the ring indicating that local pollution processes are absent.  

Although the underlying stellar population of the host galaxy is detected in the spectrum of the
central knot (knot C), the similarity of the star formation histories in the four knots, as deduced from the fitting of SSPs, as well as the  derived metallicities, point to a common chemical evolution of the polar ring.

We calculated the SFR in the different knots studied from the H$\alpha$ luminosities. The combined SFR for the ring amounts to half of the integrated SFR for this galaxy reported by previous authors. The average SFR is also higher than the average reported value for HII regions in polar rings by more than an order of magnitude.

Finally, for the differential velocity, we used the wavelength position of H$\alpha$
and [S{\sc ii}] to measure an assymmetric rotation of the ring with a mean value
of 85 km/s at an optical radius of 20''. This gives a dynamical mass of (2.8 $\pm$ 0.2) $\cdot$ 10$^9$ M$_\odot$ and an M/L$_B$ ratio of 3.9, close to the value reported previously by Reshetnikov \& Combes (1994).
The kinematics of the ring is significantly affected by the expanding bubbles of ionised gas, which
in the case of knot C reaches 60 km/s.

\begin{acknowledgements}
This work has been partially supported by projects AYA2004-02703, AYA2007-67965-C03-02, AYA2007-67965-C03-03, and AYA2007-64712
 of the Spanish National Plan for Astronomy and Astrophysics and the CNRS-INSU and its Programmes Nationaux de Galaxies et de Cosmologie (France). Also, partial support from the Comunidad de Madridunder grant S0505/ESP000237 (ASTROCAM) is acknowledged. EPM acknowledges his financial support from the {\em Fundaci\'on Espa\~nola para la Ciencia y la Tecnolog\'\i a} during his two-year postdoctoral position in the LATT/OMP.

The WHT is operated on the island of La Palma by the ING in the Spanish Observatorio del Roque de los
Muchachos of the Instituto de Astrof\'\i sica de Canarias. We thank the Spanish allocation committee (CAT)
for awarding observing time.

We would like to thank to Armando Gil de Paz, for allowing the study of B, R, and H$\alpha$ images of IIZw71 in the Palomar/Las Campanas Atlas of Blue Compact Galaxies
and to Guillermo F. H\"agele for very interesting discussions and comments that have helped to improve this work. 
We thank Roberto Cid Fernandes and
the people of the Starlight Project Team (UFSC, Brazil), for making the STARLIGHT code available. We also thank the anonymous referee
for his/her constructive comments.
\end{acknowledgements}


\begin{thebibliography}{}


\bibitem[Alloin et al.(1979)]{1979A&A....78..200A} Alloin, D., 
Collin-Souffrin, S., Joly, M., \& Vigroux, L.\ 1979, \aap, 78, 200 
\bibitem{} Barker, T. 1980, ApJ, 240, 99.
\bibitem[Bournaud \& Combes(2003)]{2003A&A...401..817B} Bournaud, F., \& 
Combes, F.\ 2003, \aap, 401, 817 
\bibitem[Bruzual \& Charlot(2003)]{2003MNRAS.344.1000B} Bruzual, G., \& 
Charlot, S.\ 2003, \mnras, 344, 1000 
\bibitem[Cair{\'o}s et al.(2001)]{2001ApJS..133..321C} Cair{\'o}s, L.~M., 
V{\'{\i}}lchez, J.~M., Gonz{\'a}lez P{\'e}rez, J.~N., Iglesias-P{\'a}ramo, 
J., \& Caon, N.\ 2001a, \apjs, 133, 321 
\bibitem[Cair{\'o}s et al.(2001)]{2001ApJS..136..393C} Cair{\'o}s, L.~M., 
Caon, N., V{\'{\i}}lchez, J.~M., Gonz{\'a}lez-P{\'e}rez, J.~N., \& 
Mu{\~n}oz-Tu{\~n}{\'o}n, C.\ 2001b, \apjs, 136, 393 
\bibitem[Cardelli et al.(1989)]{1989ApJ...345..245C} Cardelli, J.~A., Clayton, G.~C., \& Mathis, J.~S.\ 1989, \apj, 345, 245 
\bibitem[Chabrier(2003)]{2003PASP..115..763C} Chabrier, G.\ 2003, \pasp, 
115, 763 
\bibitem[Cid Fernandes et al.(2004)]{2004MNRAS.355..273C} Cid Fernandes, 
R., Gu, Q., Melnick, J., Terlevich, E., Terlevich, R., Kunth, D., Rodrigues 
Lacerda, R., \& Joguet, B.\ 2004, \mnras, 355, 273 
\bibitem{} Cid-Fernandes, R., Mateus, A., Sodr\'e, L., Stasi{\'n}ska, G. \& Gomes, J.M. 2005, MNRAS, 358, 363.
\bibitem{} Cox, A.L., Sparke, L.S., Watson, A.M., von Moorsel, G. 2001, AJ, 121, 692.
\bibitem[Denicol{\'o} et al.(2002)]{2002MNRAS.330...69D} Denicol{\'o}, G., 
Terlevich, R., \& Terlevich, E.\ 2002, MNRAS, 330, 69 
\bibitem{} De Robertis, M.M., Dufour, R.J. \& Hunt, R.W. 1987, JRASC, 81, 195
\bibitem[D\'\i az(1988)]{1988MNRAS.231...57D} D\'\i az, A.~I.\ 1988, \mnras, 231, 57 
\bibitem[D{\'{\i}}az \& P{\'e}rez-Montero(2000)] {} D{\'{\i}}az, A.~I., \& P{\'e}rez-Montero, E.\ 2000, MNRAS, 312, 130 
\bibitem[D{\'{\i}}az et al. (2000)]{} D\'\i az, A. I., Castellanos, M., Terlevich, E. \& Garc\'\i a-Vargas, M.L.\ 2000, MNRAS, 318, 462 
\bibitem{} Filippenko, A. V. 1982, PASP, 94, 715
\bibitem{} Garnett, D.R. 1992, AJ, 103, 1330.
\bibitem{} Gil de Paz, A., Madore, B.F. \& Pevunova, O. 2003, ApJSS, 147, 29.
\bibitem[H{\"a}gele et al.(2006)]{2006MNRAS.372..293H} H{\"a}gele, G.~F., 
P{\'e}rez-Montero, E., D{\'{\i}}az, {\'A}.~I., Terlevich, E., \& Terlevich, 
R.\ 2006, \mnras, 372, 293 
\bibitem[Izotov et al.(2004)]{2004A&A...415...87I} Izotov, Y.~I., 
Stasi{\'n}ska, G., Guseva, N.~G., \& Thuan, T.~X.\ 2004, \aap, 415, 87 
\bibitem{}  Kauffmann, G., White, S.D.M. \& Guiderdoni, B. 1993, MNRAS, 264, 201
\bibitem[Kennicutt(1998)]{1998ARA&A..36..189K} Kennicutt, R.~C., Jr.\ 1998, 
\araa, 36, 189 
\bibitem{} Kewley, L., J., Jansen, R.A. \& Geller, M.J. 2005, PASP, 117, 227
\bibitem[Le Borgne et al.(2003)]{2003A&A...402..433L} Le Borgne, J.-F., et 
al.\ 2003, \aap, 402, 433 
\bibitem[Mateus et al.(2006)]{2006MNRAS.370..721M} Mateus, A., Sodr{\'e}, 
L., Cid Fernandes, R., Stasi{\'n}ska, G., Schoenell, W., \& Gomes, J.~M.\ 
2006, \mnras, 
\bibitem[McGaugh(1991)]{1991ApJ...380..140M} McGaugh, S.~S.\ 1991, \apj, 
380, 140 
\bibitem[Pagel et al.(1979)]{1979MNRAS.189...95P} Pagel, B.~E.~J., Edmunds, 
M.~G., Blackwell, D.~E., Chun, M.~S., \& Smith, G.\ 1979, \mnras, 189, 95 
\bibitem{} P\'erez-Montero, E. \& D\'iaz, A.I. 2003, MNRAS, 346, 105.
\bibitem{} P\'erez-Montero, E. \& D\'iaz, A.I. 2005, MNRAS, 361, 1063.
\bibitem[P{\'e}rez-Montero et al.(2006)]{2006A&A...449..193P} 
P{\'e}rez-Montero, E., D{\'{\i}}az, A.~I., V{\'{\i}}lchez, J.~M., \& 
Kehrig, C.\ 2006, \aap, 449, 193 
\bibitem[\protect\citeauthoryear{P{\'e}rez-Montero et 
al.}{2007}]{2007MNRAS.381..125P} P{\'e}rez-Montero E., H{\"a}gele G.~F., 
Contini T., D{\'{\i}}az {\'A}.~I., 2007, MNRAS, 381, 125 
\bibitem[Pettini \& Pagel(2004)]{2004MNRAS.348L..59P} Pettini, M., \& 
Pagel, B.~E.~J.\ 2004, \mnras, 348, L59 
\bibitem{} Reshetnikov, V.P. \& Combes, F. 1994, A\&A, 291, 57.
\bibitem[Shaw \& Dufour(1995)]{1995PASP..107..896S} Shaw, R.~A., \&
Dufour,
R.~J.\ 1995, PASP, 107, 896
\bibitem[Shi et al.(2005)]{2005A&A...437..849S} Shi, F., Kong, X., Li, C., 
\& Cheng, F.~Z.\ 2005, A\&A, 437, 849 
\bibitem{} Stasi\'nska, G. 1978, A\&A, 66, 257.
\bibitem[Stasi{\'n}ska(2006)]{2006A&A...454L.127S} Stasi{\'n}ska, G.\ 2006, 
\aap, 454, L127 
\bibitem[Storchi-Bergmann et al.(1994)]{1994ApJ...429..572S} 
Storchi-Bergmann, T., Calzetti, D., \& Kinney, A.~L.\ 1994, ApJ, 429, 572 
\bibitem[Storey \& Hummer(1995)]{1995MNRAS.272...41S} Storey, P.~J., \& 
Hummer, D.~G.\ 1995, \mnras, 272, 41 
\bibitem{} V\'{\i}lchez, J.M. \& Esteban, C. 1996, MNRAS, 280, 720.
\bibitem{}  Whitmore, B.C, Lucas, R.A, McElroy, D.B., Steiman-Cameron, T.Y., Sackett, 
P.D. \& Olling, R.P. 1990, AJ, 100, 1489.




\end{thebibliography}
\end{document}